%% file: template.tex
\RequirePackage{fix-cm}
\documentclass[twocolumn]{svjour3}          
\smartqed  
\usepackage{graphicx}
\usepackage[colorinlistoftodos,textsize=scriptsize]{todonotes}
\newboolean{uprightparticles}
\setboolean{uprightparticles}{false} 

\input{lhcb-symbols-def} 

\usepackage{xspace} 
\usepackage{upgreek}
\usepackage{amsmath}
\usepackage{hyperref}
\usepackage{lineno}

\addcontentsline{toc}{section}{References}

\usepackage{cite} 
\usepackage{mciteplus}

\begin{document}

\title{Charm \CPV: observation and prospects}

\author{  Miroslav Saur      \and
        Fu-Sheng Yu  
}


\institute{Miroslav Saur \at
              School of Physical Sciences, University of Chinese Academy of Sciences, Beijing, 100000, China \\
              \email{miroslav.saur@cern.ch}
                \and
                Fu-Sheng Yu \at
              School of Nuclear Science and Technology, Lanzhou university, Lanzhou, 730000, China \\
              \email{yufsh@lzu.edu.cn}           
}


\maketitle


\section{Introduction}\label{intro}

In physics, the symmetries and their violation always provide deep insights into the Nature.  
The parity ($P$) symmetry represents the system is unchanged under the space reflection. The violation of parity, firstly proposed by Lee and Yang and subsequently discovered in 1956, plays the key role in the understanding of the weak interaction which is one of the four basic forces of nature. 
The charge ($C$) symmetry describes a property between particles and their anti-particles. 
The violation of the combined charge-parity (\CP) symmetry was unexpectedly observed in kaon meson decays in 1964. 
The $C$ and \CP violation (\CPV) are required to explore why there are much more matter than anti-matter in the Universe.

The explanation of \CPV was proposed by Kobayashi and Maskawa (KM) in 1973 by introducing three generations of quarks, or say six quarks, whereas only three quarks were established at the time. All the six quarks were found in the following twenty years.
This theory was finally manifested by the observation of \CPV in the bottom-quark meson system in 2001. 
The measured amount of \CPV in the Standard Model (SM) of particle physics is about ten orders of magnitude smaller than required by the matter-antimatter asymmetry in the Universe. 
Therefore, it is important to search for new sources of \CPV beyond the SM (BSM).
The KM mechanism also predicts the existence of \CPV in the charm-quark system which, however, had never been discovered with a lot of efforts during the past decade. 
The LHCb collaboration eventually observed the charm \CPV in 2019 via measuring the difference of \CP asymmetries of $D^0\to K^+K^-$ and $D^0\to\pi^+\pi^-$ with the result of $(1.54\pm0.29)\times10^{-3}$ \cite{LHCb-PAPER-2019-006}, with the significance of 5.3$\sigma$. 
After the establishment of \CPV in the strange- and bottom-quark systems, the observation of charm \CPV is a milestone of particle physics. 

\section{LHCb and recent measurement}
Large Hadron Collider beauty experiment (\lhcb) on Large Hadron Collider (\lhc) is a dedicated heavy-flavour (particles containing \cquark and \bquark quarks) experiment
with a special focus on \CPV measurements. Being a single-arm forward spectrometer with excellent vertex, interaction point and momentum resolution in combination with high efficient particle identification systems and large \ccbar cross-section, \lhcb can study charm physics, especially possible \CP violating processes, with higher precision than previous dedicated $B$-factory experiments.

In the time period from 2011 to 2018, \lhcb has collected 9 fb$^{-1}$ of data, roughly corresponding to the sample of decays of $10^{10}$ \Dz whose components are a charm quark and an anti-up quark. Charmed mesons can be produced as a direct result of proton-proton collisions (prompt production) or via weak decays of \bquark-hadrons (semileptonic productions). In the case of studies using \Dz mesons, prompt production is in fact a strong decay \mbox{$\D^*(2010)^+ \rightarrow \Dz \pip$} and charge conjugated decay as well. Usage of this decays allows to determine exact charm charge of \Dz meson according to the charge of bachelor pion. Semileptonic process are then defined by the weak decay \mbox{$\Bzb \rightarrow \Dz \mup \neumb X$} and charge conjugated, where $X$ stands for any allowed additional particles. In this case charm charge of \Dz meson is determined by the charge of muon. 

Recently reported observation of \CPV in the Charm sector by \lhcb is based on the new Run 2 data analysis and subsequent combination of the obtained results with the previous measurements from Run 1 \cite{LHCb-PAPER-2014-013,LHCb-PAPER-2015-055}. The new analysis is based on \mbox{$44~(9) \times 10^6$} 
and \mbox{$14~(3) \times 10^6$} \mbox{\Dz $\rightarrow$ \Kp \Km} and \mbox{\Dz $\rightarrow$ \pip \pim} prompt  (semileptonic) decays, respectively. This data set, corresponding to \mbox{6 fb$^{-1}$,} was recorded from 2015 to 2018 at the collision energy \mbox{13~TeV}.


Time dependent \CP asymmetry of \Dz decays is given by
\begin{equation}
\acp(f, t) \equiv \frac{\mathrm{\Gamma}(D^0(t) \rightarrow f ) - \mathrm{\Gamma}(\Dzb(t) \rightarrow f)}{\mathrm{\Gamma}(D^0(t) \rightarrow f) + \mathrm{\Gamma}(\Dzb(t) \rightarrow f)},
\label{eq:acp_time}
\end{equation}
where \textit{f} is a final state and \CP eigenstate, in the case of reported analysis final state is \Kp\Km or \pip\pim and \Dzb is the anti-particle of \Dz.
This asymmetry can be also written as the combination of direct and indirect \CP asymmetry effect: 
   $ \acp(f) \approx a_{\CP}^{\mathrm{dir}}(f) - \frac{\langle t (f) \rangle}{\tau (\Dz)} \agamma (f)$,
where $\langle t (f) \rangle$ denotes the mean decay time of \Dz $\rightarrow~f$ influenced by the experimental efficiency, $a_{\CP}^{\mathrm{dir}}(f)$ is the direct \CP asymmetry, $\tau$(\Dz) the \Dz lifetime and \agamma the asymmetry between the \Dz $\rightarrow f$ and \Dzb $\rightarrow f$ effective decay widths.

\begin{figure}[!]
	\includegraphics[width=0.50\textwidth]{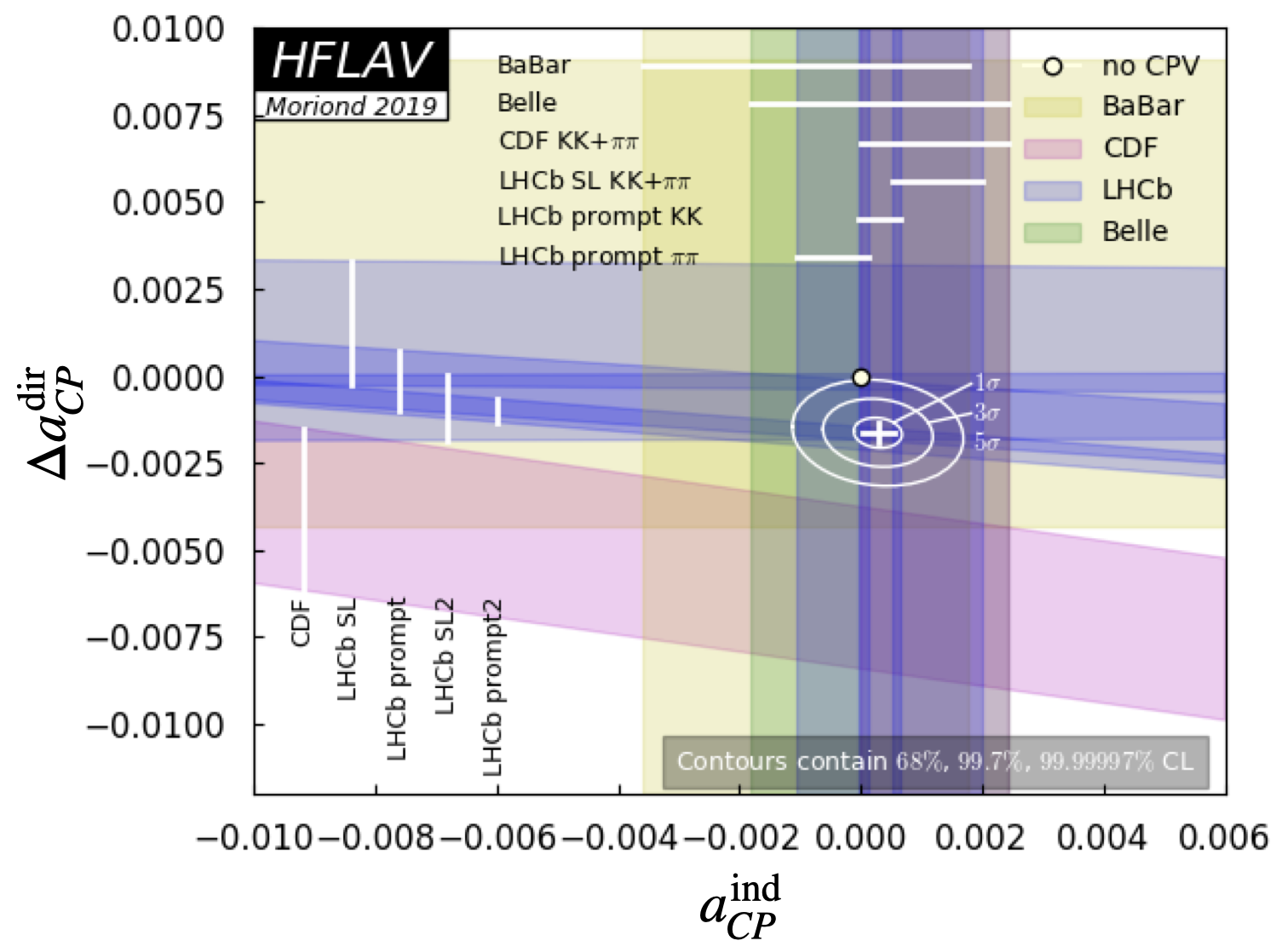}
	\caption{HFLAV fit of direct \CPV parameter $\Delta a^{\mathrm{dir}}_{\CP}$ and indirect \CPV parameter $a_{\CP}^{\mathrm{ind}}$ updated with the reported LHCb measurement. Reproduced from Ref. \cite{HFLAV16}.}
	\label{fig:hflav_fit}
\end{figure}

However,the  \acp values, as defined above are not accessible directly by the experimental methods and must be extracted from the data. Directly measurable value is the difference between raw yields, $A_{\rm raw}$, of \dzkpkm and \dzbkpkm decays or between \dzpippim and \dzbpippim, respectively. $A_{\rm raw}$ can be very well approximated, up to the order $\mathcal{O}(10^{-6})$, as linear combination of physical \CP asymmetry \acp, detection asymmetry of \Dz which is equal to zero due to charge conjugated final states, mother particle production asymmetry and detection asymmetry of tagging particle. These detection and production asymmetries are cancelled by equalising kinematics between \Kp \Km and \pip \pim decay modes and then taking a difference. This equalisation is done in three dimension of kinematic variables simultaneously after the removal of phase space regions with large intrinsic asymmetries due to the \lhcb detector geometry. Final experimental formula is then written as following
\begin{align}
\dacp &\equiv \acp (D^0\to\Kp \Km) - \acp(D^0\to\pip \pim)  \nonumber\\
&=A_{\rm raw}^{\rm equalised}(\Kp \Km) - A_{\rm raw}^{\rm equalised}(\pip \pim).
\label{eq:dacp}
\end{align}

The difference of \CP asymmetries in $D^0\to K^+K^-$ and $D^0\to \pi^+\pi^-$ are finally measured by LHCb as 
$\dacp^{\rm prompt} = [-18.2 \pm 3.2 \stat \pm 0.9 \syst] \times 10^{-4}$, $\dacp^{\rm semileptonic} = [-9 \pm 8 \stat \pm 5 \syst] \times 10^{-4}$ \cite{LHCb-PAPER-2019-006}. 
By combing both these results with the previous LHCb measurements with the Run I data of 3 fb$^{-1}$ \cite{LHCb-PAPER-2014-013,LHCb-PAPER-2015-055}, it can be obtained that
\begin{equation}
\begin{split}
\dacp^{\rm combined} = (-15.4 \pm 2.9) \times 10^{-4},
\label{eq:results_combined}
\end{split}
\end{equation}
where the uncertainty includes statistical and systematic contributions. This result deviates from zero \CP asymmetry hypothesis on \mbox{5.3$\sigma$} level. This is the first observation of \CP violation in the charm sector. 

With the \lhcb average of \agamma \cite{HFLAV16}, the direct \CP asymmetry can then be obtained as \\
\mbox{$\Delta a_{\CP}^{\mathrm{dir}} = (-15.7 \pm 2.9) \times 10^{-4}$},
which shows the sensitivity of \dacp to the direct \CPV. 
Finally, the combined fit of the direct and indirect \CP asymmetries by the Heavy Flavour Averaging Groups (HFLAV) is shown in Fig. \ref{fig:hflav_fit}. The current world average result excludes the no-\CPV hypothesis on the level of \mbox{5.44$\sigma$}.


\section{Theoretical explanations and implications}

In theory, \CPV in $D^0\to K^+K^-$ and $\pi^+\pi^-$ results from the interference between the tree and penguin amplitudes of charm decays.
It is difficult to calculate in the first-principle QCD methods due to the large non-perturbative contributions at the charm scale. Therefore, the order of magnitude of predictions on the charm \CPV is meaningful.

\begin{figure}[!]
	\includegraphics[width=0.45\textwidth]{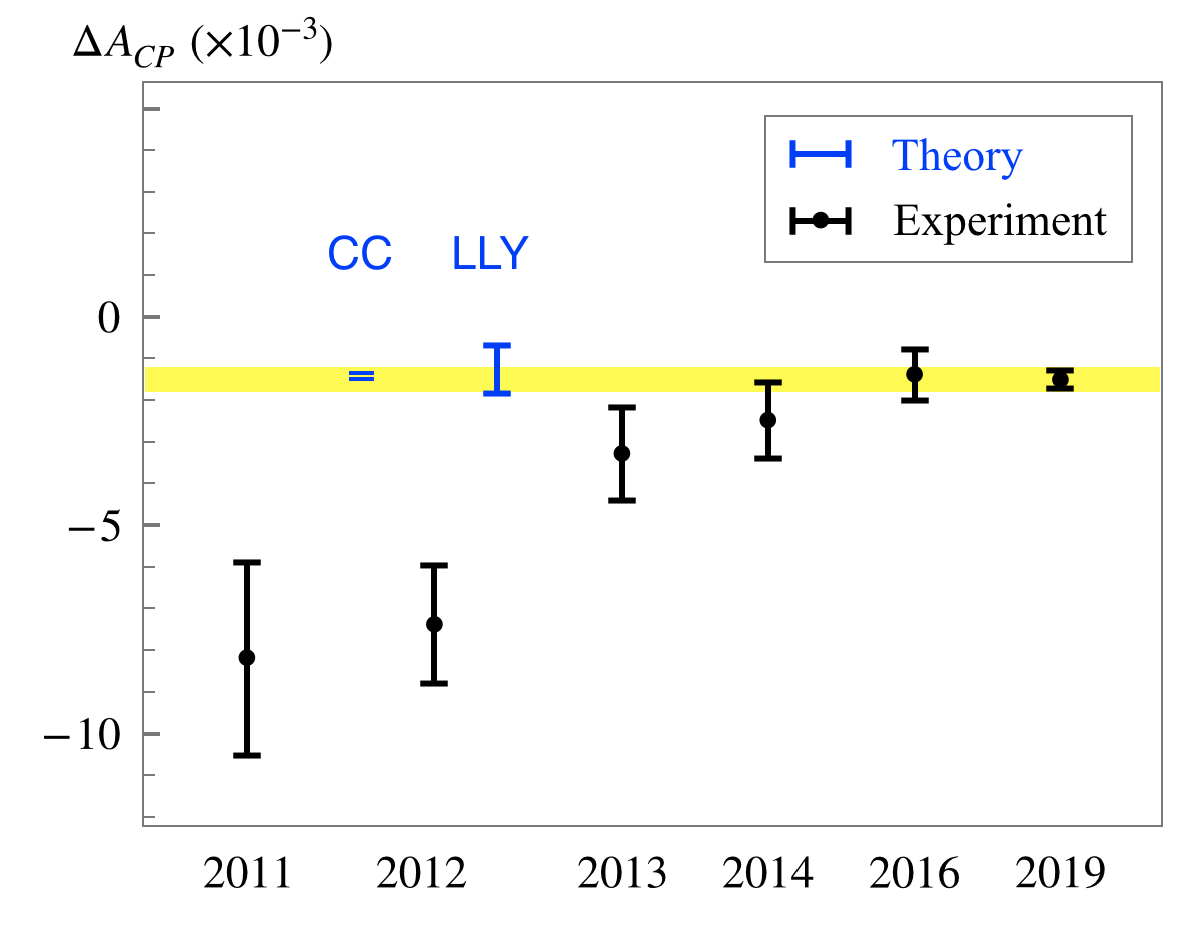}
	\caption{Comparison between experimental measurements (in black) and theoretical predictions (in blue) on \dacp in year. Experimental results are corresponding to the world-average values for specific year as calculated by the HFLAV~\cite{HFLAV16}. The theoretical approaches of CC and LLY are explained in text. The yellow band is the most recent experimental result for comparison.}\label{fig:CPVyear}
\end{figure}

Before 2019, several orders of magnitude of charm \CPV have been predicted in literatures, ranging from $10^{-4}$ to $10^{-2}$. If persisting in using the perturbative QCD, \CPV in charm decays is naively expected as $A_{CP}\sim {\alpha_s(\mu_c)\over\pi}{|V_{ub}V_{cb}^*|\over |V_{us}V_{cs}^*|}=\mathcal{O}(10^{-4})$. On the contrary, If taking the unknown non-perturbative contributions to be arbitrarily large, the charm \CPV could be as large as $10^{-2}$, to be consistent with the experimental results in 2011 when \dacp was measured to be $(-0.82\pm0.24)\%$ by \lhcb \cite{Aaij:2011in}. Due to the limit of space of this article, relevant references can be seen in \cite{FSY_implication:2019hho}. The most interesting thing is that only two papers, written by Cheng and Chiang (CC)  \cite{Cheng:2012xb} and Li, L\"u and Yu (LLY)  \cite{Li:2012cfa}, quantitatively predicted \dacp at the order of $10^{-3}$ before the observation. They are much smaller than the experimental measurements in 2011 and 2012, but manifested by the recent LHCb result. The comparison between the experimental measurements and the theoretical predictions by CC and LLY are shown in Fig. \ref{fig:CPVyear}.

To predict the charm \CPV, one should reliably obtain the tree amplitudes first, to understand the dynamics at the charm scale, and then calculate the penguin amplitudes reasonably. 
Including all the strong-interaction effects, especially the non-perturbative contributions, the topological diagrammatic approach works well for hadronic charm-meson decays by extracting the tree amplitudes from the data of branching fractions \cite{Cheng:2012wr,Cheng:2012xb,Li:2012cfa}. 
CC pointed out that the upper bound of \dacp in the SM is $-0.25\%$ \cite{Cheng:2012wr} which is more than $2\sigma$ away from the experimental result at that time. Later on, they assumed that the penguin-exchange diagram is identical to the $W$-exchange diagram, $PE=E$, so that \dacp$= (-1.39 \pm 0.04) \times 10^{-3}~\mathrm{or}~(-1.51 \pm 0.04) \times 10^{-3}$ \cite{Cheng:2012xb}, and updated with similar results in \cite{Cheng:2019ggx}. 
To give a reasonable uncertainty of the CC approach, considering the possible difference between $PE$ and $E$, we take $PE$ ranging from $E/2$ to $2E$. 
Under the factorization hypothesis, LLY proposed the factorization-assisted topological-amplitude approach which relates the penguin amplitudes to the tree amplitudes without any additional free parameters. Considering the uncertainties of input parameters, it is predicted that \mbox{\dacp$= (-0.57 \sim -1.87) \times 10^{-3}$} \cite{Li:2012cfa}, which is consistent with the latest result by the LHCb measurement.

After the observation of charm \CPV by LHCb in 2019, new explanations are explored either in the SM or in the BSM. In the SM picture, the measured result of \dacp can be explained by the non-perturbative final-state-interaction contributions from the rescattering effects \cite{Grossman:2019xcj} or the near-by resonant effects \cite{Soni:2019xko}. 
Alternatively, given that the SM contribution to the charm \CPV is very small based on the heavy-quark expansion and the perturbative QCD, the observed \dacp are explored by the BSM explanations, such as the flavour-violating $Z'$ model, the two-Higgs-doublet model, and vector-like quark models \cite{Lenz_CPV_BSM:2019fdb,Dery_implication:2019ysp,Calibbi:2019bay}.

Due to the non-perturbative physics at the charm scale, it is difficult to use the observed CPV to reliably search for new physics. On the contrary, the study of charm CPV could be used to test and understand the non-perturbative behaviour of the SM. The additional progress in theory and also more precise experimental measurements are needed.

\section{Impact and prospect for the future}

Although the combination of \dzkpkm and \dzpippim was expected to be one of the best probes of \CPV in charm, many other studies are possible or even already ongoing, such as $\Dz \rightarrow \KS \KS$ and $\Dp \rightarrow \Kp \Km \pip$ \cite{FSY_implication:2019hho}, to test and understand the dynamics of charm decay and to search for new physics beyond the SM. 
Due to their generally rich decays structure multibody decays offer additional interesting possibility how to look for \CPV signatures. However, at the same time such a studies generally require more complicated analysis methods and higher recorded luminosity. 
Another interesting measurement is being performed to investigate a novel effect of \CPV in charmed meson decaying into \KS, which comes from mother decay and daughter mixing with predicted values reaching the available experimental sensitivity \cite{Yu_CPV:2017oky}.

The \lhcb detector is currently going through the substantial upgrade 
which will allow to record data with 5 times higher luminosity than during the years 2015-2018. This, in combination with the new full software trigger, a crucial point for the charm physics, will allow \lhcb to achieve an unprecedented precision in the heavy flavour sector \CPV measurements. 
This opens a door to measure possible \CPV effects in rare decays, \eg radiative and semi-leptonic decays. Another dedicated heavy-flavour experiment is \belletwo, which started taking data in 2019, from which contributions to \CPV measurements are expected, especially results from the decays with neutral particles in the final states.
Another substantial upgrade of the \lhcb is planned for the time period after 2030, with additional ten fold increase of luminosity. The \lhcb is currently expected to be the only dedicated heavy-flavour experiment taking data during that time period. \mbox{Table \ref{tab:prospect}}~summarises current and expected future trigger-level yields and sensitivity in promptly produced \dzkpkm and \dzpippim decays.

In summary,  the first experimental observation of \CP violation in the charm sector was done with an amazing sensitivity obtained by \lhcb in 2019. This is a milestone in the high energy physics. The result is consistent with the theoretical predictions by CC and LLY. It is expected that more precise measurements and more theoretical studies in the near future will help us to deeply understand the dynamics at the charm scale and to explore the new physics effects.

\begin{table}
	\caption{Overview of the recorded and predicted values for promptly produced $\Dz \rightarrow \Kp \Km$ and $\Dz \rightarrow \pip \pim$ yields during the different data-taking periods at \lhcb. All values are corresponding to the trigger-level yields within the \lhcb acceptance. Last column shows expected precision of the \dacp measurements with the corresponding yields. Reproduced from Ref.~\cite{LHCb-PII-Physics}.}
	\label{tab:prospect}       
	\begin{tabular}{llll}
		\hline
		Sample [fb$^{-1}$] & Yield & yield & $\sigma (\dacp)$\\
		& $\Dz \rightarrow \Kp \Km$ & $\Dz \rightarrow \pip \pim$ & [\%]\\
		\noalign{\smallskip}\hline\noalign{\smallskip}
		Run 1-2 (9) & $52 \times 10^6$ & $17 \times 10^6$ & 0.03\\
		Run 1-3 (23) & $280 \times 10^6$ & $94 \times 10^6$ & 0.013\\
		Run 1-4 (50) & $1 \times 10^9$ & $305 \times 10^6$ & 0.01\\
		Run 1-5 (300) & $4.9 \times 10^9$ & $1.6 \times 10^9$ & 0.003\\
		\hline
	\end{tabular}
\end{table}

\section{Acknowledgement}
This work is partially supported by the National Natural Science Foundation of China under \\
Grants No.U1732101 and 11975112, by Gansu Natural Science Fund under grant No.18JR3RA265, by the Fundamental Research Funds for the Central Universities under Grant No. lzujbky-2019-55 and by the University of Chinese Academy of Sciences scholarship for International students. 
\bibliographystyle{LHCb}
\bibliography{local}   
\end{document}

%% file: lhcb-symbols-def.tex
\usepackage{xspace} 
\usepackage{upgreek}

\def\lhcb   {\mbox{LHCb}\xspace}

\def\belletwo {\mbox{Belle~II}\xspace}

\def\lhc    {\mbox{LHC}\xspace}

\def\MagUp {\mbox{\em Mag\kern -0.05em Up}\xspace}

\ifthenelse{\boolean{uprightparticles}}%
{

 \def\Pmu         {\ensuremath{\upmu}\xspace}                 
 \def\Pnu         {\ensuremath{\upnu}\xspace}                 
                  
 \def\Ppi         {\ensuremath{\uppi}\xspace}

 \def\PDelta      {\ensuremath{\Delta}\xspace}                 
 \def\PXi         {\ensuremath{\Xi}\xspace}                 
 \def\PLambda     {\ensuremath{\Lambda}\xspace}                 
 \def\PSigma      {\ensuremath{\Sigma}\xspace}                 
 \def\POmega      {\ensuremath{\Omega}\xspace}                 
 \def\PUpsilon    {\ensuremath{\Upsilon}\xspace}

 \def\PB      {\ensuremath{\mathrm{B}}\xspace}                 
                  
 \def\PD      {\ensuremath{\mathrm{D}}\xspace}

 \def\PK      {\ensuremath{\mathrm{K}}\xspace}

 \def\Pb      {\ensuremath{\mathrm{b}}\xspace}                 
 \def\Pc      {\ensuremath{\mathrm{c}}\xspace}

 \def\Pi      {\ensuremath{\mathrm{i}}\xspace}

 \def\Ps      {\ensuremath{\mathrm{s}}\xspace}

 \def\thebaroffset{0.0em}
}
{

 \def\Pmu         {\ensuremath{\mu}\xspace}                 
 \def\Pnu         {\ensuremath{\nu}\xspace}                 
                  
 \def\Ppi         {\ensuremath{\pi}\xspace}

 \mathchardef\PDelta="7101
 \mathchardef\PXi="7104
 \mathchardef\PLambda="7103
 \mathchardef\PSigma="7106
 \mathchardef\POmega="710A
 \mathchardef\PUpsilon="7107
                  
 \def\PB      {\ensuremath{B}\xspace}                 
                  
 \def\PD      {\ensuremath{D}\xspace}

 \def\PK      {\ensuremath{K}\xspace}

 \def\Pb      {\ensuremath{b}\xspace}                 
 \def\Pc      {\ensuremath{c}\xspace}

 \def\Pi      {\ensuremath{i}\xspace}

 \def\Ps      {\ensuremath{s}\xspace}

 \def\thebaroffset{0.18em}
}
\newcommand{\offsetoverline}[2][\thebaroffset]{\kern #1\overline{\kern -#1 #2}}%

\makeatletter
\ifcase \@ptsize \relax
  \newcommand{\miniscule}{\@setfontsize\miniscule{4}{5}}
\or
  \newcommand{\miniscule}{\@setfontsize\miniscule{5}{6}}
\or
  \newcommand{\miniscule}{\@setfontsize\miniscule{5}{6}}
\fi
\makeatother

\DeclareRobustCommand{\optbar}[1]{\shortstack{{\miniscule (\rule[.5ex]{1.25em}{.18mm})}
  \\ [-.7ex] $#1$}}

\def\mup        {{\ensuremath{\Pmu^+}}\xspace}

\def\neub       {{\ensuremath{\overline{\Pnu}}}\xspace}

\def\neumb      {{\ensuremath{\neub_\mu}}\xspace}

\def\squark    {{\ensuremath{\Ps}}\xspace}

\def\cquark    {{\ensuremath{\Pc}}\xspace}
\def\cquarkbar {{\ensuremath{\overline \cquark}}\xspace}
\def\ccbar     {{\ensuremath{\cquark\cquarkbar}}\xspace}
\def\bquark    {{\ensuremath{\Pb}}\xspace}

\def\pion   {{\ensuremath{\Ppi}}\xspace}

\def\pip    {{\ensuremath{\pion^+}}\xspace}
\def\pim    {{\ensuremath{\pion^-}}\xspace}

\def\kaon    {{\ensuremath{\PK}}\xspace}

\def\KorKbar {\kern \thebaroffset\optbar{\kern -\thebaroffset \PK}{}\xspace}

\def\Kp      {{\ensuremath{\kaon^+}}\xspace}
\def\Km      {{\ensuremath{\kaon^-}}\xspace}

\def\KS      {{\ensuremath{\kaon^0_{\mathrm{S}}}}\xspace}

\def\Dbar    {{\ensuremath{\offsetoverline{\PD}}}\xspace}
\def\D       {{\ensuremath{\PD}}\xspace}

\def\DorDbar {\kern \thebaroffset\optbar{\kern -\thebaroffset \PD}\xspace}
\def\Dz      {{\ensuremath{\D^0}}\xspace}
\def\Dzb     {{\ensuremath{\Dbar{}^0}}\xspace}
\def\Dp      {{\ensuremath{\D^+}}\xspace}

\def\B       {{\ensuremath{\PB}}\xspace}
\def\Bbar    {{\ensuremath{\offsetoverline{\PB}}}\xspace}

\def\BorBbar {\kern \thebaroffset\optbar{\kern -\thebaroffset \PB}\xspace}

\def\Bzb     {{\ensuremath{\Bbar{}^0}}\xspace}
\def\Bd      {{\ensuremath{\B^0}}\xspace}

\def\BdorBdbar {\kern \thebaroffset\optbar{\kern -\thebaroffset \Bd}\xspace}

\def\Bs      {{\ensuremath{\B^0_\squark}}\xspace}

\def\BsorBsbar {\kern \thebaroffset\optbar{\kern -\thebaroffset \Bs}\xspace}

\def\Y#1S{\ensuremath{\PUpsilon{(#1S)}}\xspace}

\def\LorLbar     {\kern \thebaroffset\optbar{\kern -\thebaroffset \PLambda}\xspace}

\newcommand{\decay}[2]{\ensuremath{#1\!\to #2}\xspace} 

\def\to                 {\ensuremath{\rightarrow}\xspace}

\def\CP                {{\ensuremath{C\!P}}\xspace}

\def\CPV               {{\ensuremath{C\!PV}}\xspace}

\def\acp		{\ensuremath{A_{CP}}\xspace}
\def\dacp		{\ensuremath{\mathrm{\Delta} A_{CP}}\xspace}

\def\AT#1     {\ensuremath{A_{\mathrm{T}}^{#1}}\xspace}           

\def\C#1      {\ensuremath{\mathcal{C}_{#1}}\xspace}                       
\def\Cp#1     {\ensuremath{\mathcal{C}_{#1}^{'}}\xspace}                    
\def\Ceff#1   {\ensuremath{\mathcal{C}_{#1}^{\mathrm{(eff)}}}\xspace}        
\def\Cpeff#1  {\ensuremath{\mathcal{C}_{#1}^{'\mathrm{(eff)}}}\xspace}       
\def\Ope#1    {\ensuremath{\mathcal{O}_{#1}}\xspace}                       
\def\Opep#1   {\ensuremath{\mathcal{O}_{#1}^{'}}\xspace}                    

\def\agamma     {\ensuremath{A_{\Gamma}}\xspace}

\def\dzkpkm     {\decay{\Dz}{\Kp\Km}}
\def\dzbkpkm     {\decay{\Dzb}{\Km\Kp}}
\def\dzpippim   {\decay{\Dz}{\pip\pim}}
\def\dzbpippim   {\decay{\Dzb}{\pim\pip}}

\newcommand{\aunit}[1]{\ensuremath{\text{\,#1}}}       

\newcommand{\tev}{\aunit{Te\kern -0.1em V}\xspace}
\newcommand{\gev}{\aunit{Ge\kern -0.1em V}\xspace}
\newcommand{\mev}{\aunit{Me\kern -0.1em V}\xspace}
\newcommand{\kev}{\aunit{ke\kern -0.1em V}\xspace}
\newcommand{\ev}{\aunit{e\kern -0.1em V}\xspace}
\newcommand{\mevc}{\ensuremath{\aunit{Me\kern -0.1em V\!/}c}\xspace}
\newcommand{\gevc}{\ensuremath{\aunit{Ge\kern -0.1em V\!/}c}\xspace}
\newcommand{\mevcc}{\ensuremath{\aunit{Me\kern -0.1em V\!/}c^2}\xspace}
\newcommand{\gevcc}{\ensuremath{\aunit{Ge\kern -0.1em V\!/}c^2}\xspace}

\newcommand{\stat}{\aunit{(stat.)}\xspace}
\newcommand{\syst}{\aunit{(syst.)}\xspace}

\def\gsim{{~\raise.15em\hbox{$>$}\kern-.85em
          \lower.35em\hbox{$\sim$}~}\xspace}
\def\lsim{{~\raise.15em\hbox{$<$}\kern-.85em
          \lower.35em\hbox{$\sim$}~}\xspace}

\def\tell1  {TELL1\xspace}
\def\ukl1   {UKL1\xspace}

\newcommand{\eg}{\mbox{\itshape e.g.}\xspace}